\documentclass[%
 reprint,
nofootinbib,
 amsmath,amssymb,
 aps,
]{revtex4-1}

\usepackage{braket}
\usepackage{graphicx}
\usepackage{dcolumn}
\usepackage{bm}

\usepackage{color}

\usepackage{soul}
\usepackage[dvipsnames]{xcolor}

\begin{document}


\title{Vacuum Fluctuations and Boundary Conditions in a Global Monopole}



\author{V. S. Barroso}
\email{barrosov@ifi.unicamp.br}
\affiliation{IFGW, Universidade Estadual de Campinas, 13083-859 Campinas, S\~ao Paulo, Brazil}
\author{J. P. M. Pitelli}
\email{pitelli@ime.unicamp.br}
\affiliation{Departamento de Matem\'atica Aplicada, Universidade Estadual de Campinas,
13083-859 Campinas, S\~ao Paulo, Brazil}%



\begin{abstract}
We study the vacuum fluctuations of a massless scalar field $\hat{\Psi}$ on the background of a global monopole. Due to the nontrivial topology of the global monopole spacetime,  characterized by a solid deficit angle parametrized by $\eta^2$, we expect that $\left<\hat{\Psi}^2\right>_{\text{ren}}$ and $\left<\hat{T}_{\mu\nu}\right>_{\text{ren}}$ are nonzero and  proportional to $\eta^2$, so  that they annul in the Minkowski limit $\eta\to0$. However, due to the naked singularity at the monopole core, the evolution of the scalar field is not unique. In fact, they are in one to one correspondence with the boundary conditions which turn into self-adjoint the spatial part of the wave operator. We show that only Dirichlet boundary condition corresponds to our expectations and gives zero contribution to the vacuum fluctuations in Minkowski limit. All other boundary conditions give nonzero contributions in this limit due to the nontrivial interaction between the field and the singularity.
\end{abstract}

\maketitle


\section{\label{sec:intro}Introduction}

Grand Unified Theories (GUT) predict spontaneous symmetry breaking of matter fields during phase transitions in the early universe. As a result, topological defects might appear in the spacetime manifold, namely Cosmic Strings, Domains Walls and Global Monopoles~\cite{vilenkin2000cosmic}. Global monopoles arise when a global $O(3)$ symmetry of a triplet scalar field $\phi^a$ ($a=1,2,3$) is spontaneously broken to $U(1)$ in the Lagrangian,
\begin{equation}
    \mathcal{L}=-\frac{1}{2}g^{\mu\nu}\partial_\mu \phi^a\partial_\nu\phi^a-\frac{\lambda}{4}(\phi^a\phi^a-\eta_0^2)^2.
\end{equation}
The fields assume a ``hedgehog'' configuration,
\begin{equation}
    \phi^a=\eta_0 f(r)\frac{x^a}{r},
\end{equation}
for $x^ax^a=r^2$, with $f(r)$ vanishing as $r$ approaches $0$ and tending to $1$ for $r$ much bigger than a typical value $r_\text{c}\sim (\sqrt{\lambda}\eta_0)^{-1}$. For $r\gg r_\text{c}$, the only nonzero components of the energy-momentum tensor are
\begin{equation}
    T_t^{~t}=T_r^{~r}=\eta_0^2/r^2,
\end{equation}
which leads to a linearly divergent energy. 

Furthermore, for typical GUT scales ($\eta_0\sim 10^{16} GeV$), the energy density is extremely high, thus one might expect gravitational effects around the monopole. Accordingly, Einstein's and field equations for a spherically symmetric solution yield the global monopole spacetime metric given by
\begin{equation}
    ds^2=-B(r)dt^2+A(r)dr^2+r^2d\Omega_2,\label{eq:gmmetric0}
\end{equation}
with Schwarzschild-like coefficients~\cite{barriola89}
\begin{equation}
    B(r)=A^{-1}(r)=1-\eta^2-\frac{2GM_\text{c}}{r},
\end{equation}
where $\eta\equiv\sqrt{8\pi G}\eta_0$ and $M_\text{c}\approx-6\pi \eta_0 \lambda^{-1/2}$~\cite{harari1990}.

We can see from the metric described by \eqref{eq:gmmetric0} that the global monopole spacetime can be treated as a highly massive core centered at the origin with characteristic radius $r_\text{c}$, plus a spherically symmetric spacetime with deficit solid angle.
Far from the core, the metric \eqref{eq:gmmetric0} can be approximated by
\begin{equation}
    ds^2=-\alpha^2dt^2+\alpha^{-2}dr^2+r^2d\Omega_2,\label{eq:gmmetric}
\end{equation}
where $\alpha^2\equiv 1-\eta^2 $ and $0<\alpha<1$. The spacetime described by \eqref{eq:gmmetric} is curved and has scalar curvature $\mathcal{R}(x)=2\eta^2/r^2$. Physically acceptable values for $\eta^2$ lie around the GUT scale, which predicts $\eta^2\sim 10^{-5}$; thus, a very small curvature. Despite the Schwarzschild-like coefficients in the metric element~\eqref{eq:gmmetric0}, there is no event horizon since the monopole's mass $M_C$ is negative. Not only the metric coefficient $B(r)$ diverges at $r=0$ but also other geometrical quantities, e.g., the scalar curvature (which scales as $r^{-2}$). Therefore, another relevant feature of the global monopole spacetime is the presence of a naked singularity at $r=0$. 

We expect that fluctuations will appear on the expectation values of quantum fields due to the curvature of the spacetime. Indeed, Mazzitelli and Lousto found them in~\cite{mazzitelli91}, where they obtained that the vacuum expectation values of the field squared $\braket{\hat{\Psi}^2}$ and the energy-momentum tensor $\braket{\hat{T}_{\mu\nu}}$ both fluctuate in order $\eta^2$. However, care must be taken when studying semiclassical effects in spacetimes having naked singularities, which is the case for the global monopole. In such spacetimes, the dynamics of the fields is not uniquely defined by initial conditions, and this represents a serious difficulty in the field quantization. In a practical sense, one may not be able to solve the Klein-Gordon equation,
\begin{equation}
    (-\square_x+M^2+\xi \mathcal{R}(x))\Psi=0,\label{eq:kg}
\end{equation}
and find a complete set of positive energy eigenfunctions defined through the whole history of the spacetime; which jeopardizes any attempt of quantizing the field operator at all times in terms of positive energy modes and annihilation/creation operators.

However, Wald and Ishibashi~\cite{wald,waldishibashi} proposed a sensible dynamical description of the field by means of the imposition of boundary conditions at the spacetime singularity. This procedure allows us to consistently quantize scalar fields in the global monopole spacetimes, which, as already discussed, has a naked singularity.  In this paper we will study quantum fields on the idealized global monopole spacetime given by the metric~(\ref{eq:gmmetric}). The boundary conditions will model, somehow, the interaction of the field with the monopole core, which will be responsible for the appearance of interesting physical effects.

We organized this article as follows. In Sec.~\ref{sec:kgns}, we will show how to consistently prescribe the dynamics of massless scalar fields in spacetimes with naked singularities. In particular, we intend to briefly discuss the procedure behind it and how the prescription of boundary conditions at the spacetime singularity will help us on our goal. Considering the appropriate boundary conditions, we will find a complete set of eigenfunctions for the Klein-Gordon operator in two cases of coupling with the curvature, namely minimal and conformal. Our choice for the coupling constant was merely for simplicity. As will be discussed later, the results found here remain valid within a certain range of positive values of the constant. For those outside this range, a more detailed discussion is required, and we leave it for future studies. With the eigenfunctions in hand, we will find the Euclidean Green's function and, from them, the quadratic vacuum expectation value of the field in Sec.~\ref{sec:gm}. We point out the contributions coming strictly from the boundary conditions, and we found its effects on the $00-$th component of the energy-momentum tensor in Sec.~\ref{sec:back}. Section~\ref{sec:concl} is devoted to our final remarks.

\section{\label{sec:kgns}Klein-Gordon equation and Naked Singularities}

As firstly proposed by Penrose in 1969~\cite{Penrose:1969pc}, gravitational collapse may always produce ``covered'' singularities, i.e., no naked singularities are allowed in nature. This statement became known as Cosmic Censorship Conjecture and was extensively developed and improved (see \cite{wald2010general}). Whether naked singularities exist or not is still an open question, but one might ask if quantum effects in General Relativity could sustain the Conjecture, or even provide some kind of ``cosmic censor''. In the absence of a well-established theory of quantum gravity, one may refer to the foundations of semiclassical gravity in order to shed some light on those topics~\cite{birrell1984quantum}. On the other hand, in spacetimes where naked singularities exist, it might not always be possible to consistently describe the evolution of quantum fields, which would inhibit any semiclassical approach. This pitfall may be overcome by the prescription of boundary conditions for the equation of motion at the spacetime singularity, i.e., imposing ``by hand'' the interaction between the fields and the classical singularity. 

\subsection{\label{sec:ngh}Non-Globally Hyperbolic static spacetimes}

In globally hyperbolic spacetimes, the dynamics arising from the Cauchy problem for well-posed initial conditions is uniquely defined at all times. Conversely, this is not true in non-globally hyperbolic spacetimes, such as those containing a naked singularity. Classical test particles following geodesics may have their trajectory interrupted at the singularity, and the future of these particles may be compromised. Thus, one may find difficulties in studying the propagation of scalar fields in the background of non-globally hyperbolic spacetimes. 

If the spacetime is also static, it can be shown that Eq.~\eqref{eq:kg} reduces to
\begin{equation}
        \frac{\partial^2\Psi}{\partial t^2} = - A \Psi,  
    \label{eq:onda}
\end{equation}
where $A := -V D^i(VD_i \phi)+M^2 V^2+\xi R V^2$ and $V^2=-\chi_\mu\chi^\mu$, where $\chi^\mu$ are the timelike Killing fields of the spacetime.
Wald argues in~\cite{wald} that $A$ can be seen as a strictly spatial differential operator acting on a Hilbert space defined over a static spatial slice $\Sigma$. Our ignorance on what happens at the singularity may be solved if we consider the domain of $A$ to be $C_0^\infty(\Sigma)$. But even though the operator $(A,C_0^\infty(\Sigma))$ is symmetric, it might not be self-adjoint. It can be shown, however, that it might admit a unique self-adjoint extension $A_E$ or an infinite set of them~\cite{reed1981functional}. 

As showed in~\cite{wald}, given initial conditions $(\Psi_0,\dot{\Psi}_0)\in C^\infty(\Sigma)\times C^\infty(\Sigma)$, the solution to Eq.~\eqref{eq:onda} for any $t\in \mathbb{R}$ is
\begin{equation}
    \Psi_t=\text{cos}(A_E^{1/2}t)\Psi_0+A_E^{-1/2}\text{sin}(A_E^{1/2}t)\dot{\Psi}_0. \label{eq:din}
\end{equation}
The dynamics associated to Eq.~\eqref{eq:din} is unique if $A$ is essentially self-adjoint, i.e., only one extension exists (Friedrichs extension).
Nonequivalent dynamics arise from Eq.~\eqref{eq:din} when the operator has infinitely many self-adjoint extensions $A_E$. To each extension, there can be associated a boundary condition at the classical singularity, labeled by the parameter $E$. Nevertheless, Wald and Ishibashi showed in~\cite{waldishibashi} that it is possible to construct a physically sensible evolution of the fields in non-globally hyperbolic spacetimes using Eq.~\eqref{eq:din}. We will discuss in Sec.~\ref{sec:sol} how this considerations can be adopted in the case of the global monopole spacetime.

\subsection{\label{sec:sol}Boundary Conditions and solutions to KG equation}

On the process of extending the domain of the operator $A$ to make it self-adjoint, one will find out the need of prescribing boundary conditions at the spacetime singularity, corresponding to each extension found. Indeed, many authors have considered these boundary conditions on the study of quantum fields propagating in non-globally hyperbolic spacetimes(see~\cite{horowitz,naked,konkowski,gurtug}). In~\cite{pitellibarroso}, the authors treated the particular case of the global monopole using Robin boundary conditions, found in~\cite{Pitelli}. We calculated the scattering pattern of massless scalar waves and, as expected, the parameter of the boundary conditions endured throughout the whole process. Direct effects appeared in the differential cross section so that one could predict which boundary condition is the one chosen by nature comparing it to experimental scattering data, if available. The parameter also influenced the stability of the spacetime, since for some particular values of it there exist bound states with divergent growth in time. 

In~\cite{pitellibarroso}, we only studied minimally coupled massless scalar fields ($\xi=0$) in the global monopole. Now,  we intend to extend the analysis to conformally coupled massless scalar fields ($\xi=1/6$) as well. To consistently do it, we must analyze Klein-Gordon equation to find the appropriate boundary conditions. Eq.~\eqref{eq:kg} can be solved in spherical coordinates under separation of variables, i.e., $\Psi(x)=e^{-i\omega t}Y_l^m(\theta,\varphi)R(r)$, and it reduces to
\begin{equation}
    -\alpha^2\left(-\frac{d^2}{d r^2} +\frac{\nu^2_\ell-1/4}{r^2}\right)u(r)=-\frac{\omega^2}{\alpha^2}u(r),\label{eq:kgmono}
\end{equation}
for $u(r)=r R(r)$ and $\nu_\ell=\frac{1}{2}\sqrt{1+\frac{4\ell(\ell+1)}{\alpha^2}+8\xi\frac{\eta^2}{\alpha^2}}$. Separation of variables simplifies the three-dimensional spatial differential operator to one depending on the radial coordinate. Thus, our problem reduces to an analog study of the well-known Calogero problem on the semiaxis in Quantum Mechanics~\cite{calogero691,calogero692,calogero71}.  In~\cite{gitman2009}, the authors study in detail the self-adjoint extensions of the Calogero operator ($A_C=-d^2/dr^2+a r^{-2}$ in $\mathcal{L}^2(\mathbb{R}^+)$) and develop its spectral analysis. They discuss the conditions on $a$ so that $A_C$ is self-adjoint or not. In our case, we identify $a\equiv \nu_\ell^2-1/4$ and two of the cases treated in~\cite{gitman2009} will be relevant, namely $a\geq 3/4$ and $-1/4<a<3/4$. We will briefly discuss each of these cases and then apply them for minimally and conformally fields to find the solutions to Eq.~\eqref{eq:kgmono}.

\paragraph{For $a\geq 3/4$} the operator $A_C$ defined over the domain $C^\infty_0(\mathbb{R}^+)$ is essentially self-adjoint, i.e., $A_C=A_C^\dagger$ and $\mathcal{D}(A_C)=\mathcal{D}(A_C^\dagger)$, and no boundary conditions are necessary. This condition requires $\nu_\ell^2\geq 1$, which implies in $\ell\geq 1$ for both cases: $\xi=0$ and $\xi=1/6$. Thus, all non spherically symmetric modes ($\ell\neq 0$) will interact trivially with the singularity at $r=0$.  The eigenfunctions of the operator $A_C$ with eigenvalues $p^2>0$ will then be
\begin{equation}
    u_{\ell, p}(r)=\sqrt{\frac{pr}{2}}J_{\nu_\ell}(pr), 
\end{equation}
where $J_\nu$ are Bessel functions of order $\nu$.
\paragraph{For $-1/4<a<3/4$} the operator $A_C$ defined over the domain $C^\infty_0(\mathbb{R}^+)$ is not self-adjoint but it admits a one-parameter $U(1)$-family of self-adjoint extensions $A_{C\beta}$ labeled by a real parameter $\beta$. This case requires the prescription of boundary conditions at $r=0$ and that $0<\nu_\ell^2< 1$. Minimally coupled fields, as well as conformally coupled ones, will only feel the effects of the boundary conditions through their spherically symmetric modes (since $0<\nu_\ell^2< 1\Rightarrow \ell=0$, for $\xi=0$ and $\xi=1/6$). This conclusion endures as long as $\xi$ is positive and much smaller than $1/\eta^2$. For some negative values of $\xi$, the inequality may not hold only for $\ell=0$, thus other modes ($\ell=1,2,\dots$) might need boundary conditions as well. The solutions must satisfy the following asymptotic boundary conditions near $r=0$
\begin{equation}
    u_\beta(r)\sim\left\{\begin{matrix}
\mathcal{N}\left[r^{1/2+\nu_0}+\beta r^{1/2-\nu_0}\right], \quad |\beta|<\infty;\\
\mathcal{N}\left[r^{1/2-\nu_0}\right], \quad |\beta|=\infty.
\end{matrix}\right.\label{eq:abcmono}
\end{equation}
The eigenvalue $p^2$ will only be positive for $\beta\geq 0$, and the associated eigenfunctions will be
\begin{equation}
    u_{0,p,\beta}(r)=\sqrt{\frac{pr}{2}}\frac{J_{\nu_0}(p r)+\gamma (\beta ,p) J_{-\nu_0}(p r)}{\sqrt{1+2 \cos (\pi  \nu_0) \gamma (\beta ,p)+\gamma (\beta ,p)^2}},\label{eq:solabc}
\end{equation}
where\footnote{The momentum parameter $p_0$ was introduced to make $\gamma(\beta, p)$ dimensionless. From now on, we will set it to one.}
\begin{equation}
    \gamma(\beta,p)=\beta\frac{\Gamma (1-\nu_0)}{\Gamma (1+\nu_0)}\left (\frac{p}{2p_0}  \right )^{2 \nu_0}.
\end{equation}
Bound states appear as one considers negatives values of $\beta$. We are interested in the continuous spectrum of the fields, hence only non-negative values of $\beta$ will be considered on what follows\footnote{We are also interested in the stable case, so that $\beta\geq 0$ (see~\cite{pitellibarroso}).}.
In the case $\xi=0$, conditions \eqref{eq:abcmono} are equivalent to Robin boundary conditions and, in accordance with~\cite{Pitelli}, the eigenfunction reduces to 
\begin{equation}
     u_{0,p,\beta}^{\xi=0}(r)=\sqrt{\frac{pr}{2}}\frac{J_{1/2}(p r)+\beta p J_{-1/2}(p r)}{\sqrt{1+(\beta p)^2}}.
\end{equation}
For $\xi=1/6$, it suffices to apply $\nu_0=\frac{1}{2}\sqrt{1+\frac{4\eta^2}{3\alpha^2}}$ on \eqref{eq:solabc}.

Upon the discussion done so far, we can establish a complete set of positive energy eigenfunctions $f_{\lambda,\ell,m}^{\xi}(x)$ to the Klein-Gordon operator ($-\square+\xi \mathcal{R}$) with eigenvalues $\lambda^2=-\omega^2/\alpha^2+p^2\alpha^2$, namely
\begin{equation}
    f_{\lambda,\ell,m}^\xi(x)=\frac{e^{-i\omega t}}{\sqrt{2\pi}r}\left\{\begin{matrix}
\frac{1}{\sqrt{4\pi}}u_{0,p,\beta}^\xi(r) \quad \text{for } \ell=0;\\ 
Y_l^m(\theta,\varphi)u_{\ell,p}^\xi(r)\quad \text{for } \ell>0.
\end{matrix}\right.\label{eq:eigenkg}
\end{equation}
A completeness relation is naturally available to them, i.e., 
\begin{equation}
        \frac{\delta^4(x,x')}{\sqrt{-g}}=\int_\lambda\sum_{\ell,m} f_{\lambda,\ell,m}(x)f^*_{\lambda,\ell,m}(x').\label{eq:complet}
\end{equation}
With these in hand, we can follow the procedure of quantization of fields. The scalar field $\Psi$ will be promoted to an operator $\hat{\Psi}$ defined as an expansion in terms of annihilation ($a(p)$) and creation ($a^\dagger(p)$) operators satisfying the canonical commutation relations. These operators will be weighted by the positive energy modes of the Klein-Gordon equation.
In the next section, we will find the vacuum expectation value of some quantities depending on the field operator.

\section{\label{sec:gm}Green's functions and $\braket{\hat{\Psi}^2}$}

Green's functions are fundamental to the computations of many quantities in Quantum Field Theory, such as scattering cross-section, decay rates, and the vacuum expectation value (v.e.v.) of the energy-momentum tensor. The latter is of particular interest for us, since it plays a relevant role in General Relativity and, notably, gives rise to semiclassical effects in gravity. As already indicated by QFT in flat spacetimes results, computations involving quadratic mean values of the field operator are expected to have divergences all over the way. In curved spacetimes, the situation is no different, and it can be even worst. That is why renormalization procedures are necessary to obtain quantities such as the v.e.v. of the energy-momentum tensor. In particular, a widely applied method is the point-splitting renormalization scheme (see~\cite{birrell1984quantum}). In~\cite{mazzitelli91}, the authors use this technique to obtain the renormalized quadratic mean value $\braket{\hat{\Psi}^2}$. To do so, they take a coincidence limit on the positions of the renormalized Green's functions, as follows
\begin{equation}
    \braket{\hat{\Psi}^2(x)}=\frac{1}{2}G^{(1)}(x,x)=iG_{F}(x,x)=G_{E}(x,x),
\end{equation}
where $G(x,x)$ indicates a formal limit $x'\rightarrow x$ on $G(x,x')$.

For convenience, Mazzitelli and Lousto follow the computation using the Euclidean Green's function, $G_E(x,x')$. For what they use the euclideanized metric of the global monopole, which is
\begin{equation}
    ds^2=\alpha^2d\tau^2+\alpha^{-2}dr^2+r^2(d\theta^2+\sin^2\theta d\varphi^2).\label{eq:gmeuclid}
\end{equation}
The Klein-Gordon operator will now have eigenvalues $\lambda^2=\omega^2/\alpha^2+p^2\alpha^2$ associated with the same eigenfunctions $f_{\lambda,\ell,m}^{\xi}(x_E)$ (under the interchange $t\rightarrow \tau$).

To evaluate the Green's function, we can use the Schwinger's integral representation
\begin{equation}
    G_E(x,x')=\int_0^\infty ds \ \exp[-s(-\square+\xi R)]\frac{\delta^4(x,x')}{\sqrt{-g}},
\end{equation}
which, using Eq.~\eqref{eq:complet}, can be expressed in terms of the eigenfunctions, as follows
\begin{multline}
    G_E(x,x')=\int_0^\infty ds \int_{-\infty}^{+\infty}d\omega \int_0^\infty dp \ e^{-s\lambda^2} \\ \times\sum_{\ell,m}f_{\lambda,\ell,m}(x)f_{\lambda,\ell,m}^*(x') 
\end{multline}
Because the boundary conditions we considered affect only spherically symmetric modes, we expect that, except by the term $\ell=0$ in the sum, the Green's function remains the same as the one found in~\cite{mazzitelli91}. Thus, it seems reasonable to separate it into two parts, one containing contributions from the boundary condition parameter ($G_E^\beta$), and the other equal to the one found by Mazzitelli and Lousto ($G_E^{ML}$). We then have
\begin{equation}
    G_E(x,x')=G_E^\beta(x,x')+G_E^{ML}(x,x'),
\end{equation}
where
\begin{multline}
     G^{ML}_E(x,x')=\int_0^\infty ds \int_{-\infty}^{+\infty}d\omega \int_0^\infty dp \ p \ e^{-s\lambda^2}
     \\\times\frac{e^{-i\omega(\tau-\tau')}}{2\pi\sqrt{r r'}}    \sum_{\ell,m}Y_\ell^m(\theta,\varphi)Y_\ell^{m*}(\theta',\varphi')J_{\nu_\ell}(pr)J_{\nu_\ell}(pr'),\label{eq:gmlsum}
\end{multline}
and\footnote{The last term subtracted in Eq.~\eqref{eq:gbetasum} was added to Eq.~\eqref{eq:gmlsum} in order to complete the sum in $\ell$ from $0$ to $\infty$.}
\begin{widetext}
\begin{multline}
    G^{\beta}_E(x,x')=\int_0^\infty ds \int_{-\infty}^{+\infty}d\omega \int_0^\infty dp \ p \ \frac{e^{-i\omega(\tau-\tau')}}{2\pi\sqrt{r r'}}\frac{e^{-s\lambda^2}}{4\pi}\\
    \times\left[\frac{(J_{\nu_0}(p r)+\gamma(\beta,p)J_{-\nu_0}(p r))(J_{\nu_0}(p r')+\gamma(\beta,p)J_{-\nu_0}(p r'))}{1+2\gamma(\beta, p)\cos \pi\nu_0+\gamma^2(\beta,p)}-J_{\nu_0}(pr)J_{\nu_0}(pr') \right].\label{eq:gbetasum}
\end{multline}
\end{widetext}
It is easy to see that the contribution $G^{\beta}_E(x,x')$ vanishes as $\beta$ goes to zero, and our the Green's function recovers identically the one from~\cite{mazzitelli91}. We will then need to compute only $G^{\beta}_E(x,x')$ in both cases of the coupling constant. Integration over $\omega$ and $s$ can be directly performed to give
\begin{widetext}
    \begin{equation}
        G^{\beta}_E(x,x')=\int_0^\infty dp \ \frac{e^{-\alpha^2p(\tau-\tau')}}{8\pi\sqrt{r r'}}\left[\frac{(J_{\nu_0}(p r)+\gamma(\beta,p)J_{-\nu_0}(p r))(J_{\nu_0}(p r')+\gamma(\beta,p)J_{-\nu_0}(p r'))}{1+2\gamma(\beta, p)\cos \pi\nu_0+\gamma^2(\beta,p)}-J_{\nu_0}(pr)J_{\nu_0}(pr') \right].\label{eq:greenbeta}
    \end{equation}
\end{widetext}  

We are interested in the effects appearing as consequence of the spacetime curvature, which, as discussed previously, scales with a factor $\eta^2\ll1$. As found in~\cite{mazzitelli91}, the fluctuations begin to appear only in first order of $\eta^2$ or higher. In comparison, we will expand our Green's function in powers of $\eta^2$ as follows
\begin{equation}
    G^{\beta}_E(x,x')=G^{\beta}_{E,0}(x,x')+\eta^2 G^{\beta}_{E,2}(x,x')+\mathcal{O}(\eta^4),\label{eq:series}
\end{equation}
so we will be able to separate which contributions appear in each order.

\subsection{\label{sec:gmmin}Minimally coupled field}

For $\xi=0$, we have $\nu_0=1/2$ and $\gamma(\beta, p)=\beta p$, and after integrating the expression above in $p$, we get to the following result
\begin{multline}
   G^{\beta}_E(\tau,\tau',r,r')= -\frac{i }{16 \pi ^2 r r'}\\ \times\left\{e^{\tilde{\aleph}_\beta} \left[-2
i \text{Ci}\left(-i\tilde{\aleph}_\beta\right)-2 \text{Si}\left(-i\tilde{\aleph}_\beta\right)+\pi \right]\right.\\
\left.-e^{\tilde{\aleph}_\beta^*} \left[2 i \text{Ci}\left(i\tilde{\aleph}_\beta^*\right)-2 \text{Si}\left(i\tilde{\aleph}_\beta^*\right)+\pi \right]\right\},\label{eq:greenminimal}
\end{multline}
where
\begin{equation}
    \tilde{\aleph}_\beta\equiv\tilde{\aleph}_\beta(\tau,\tau',r,r')=\frac{ (r+r')+i(1-\eta ^2)(\tau -\tau ')}{\beta },
\end{equation}
and $\text{Ci}$ ($\text{Si}$) is the cosine (sine) integral function. The expansion in powers of $\eta^2$ yields
\begin{multline}
   G^{\beta}_{E,0}(\tau,\tau',r,r')= -\frac{i }{16 \pi ^2 r r'}\\ \times\left\{e^{\aleph_\beta} \left[2
i \text{Ci}\left(i\aleph_\beta\right)+2 \text{Si}\left(-i\aleph_\beta\right)+\pi \right]\right.\\
\left.-e^{\aleph_\beta^*} \left[2 i \text{Ci}\left(i\aleph_\beta^*\right)-2 \text{Si}\left(i\aleph_\beta^*\right)+\pi \right]\right\},\label{eq:greenminimal0}
\end{multline}
and
\begin{multline}
   G^{\beta}_{E,2}(\tau,\tau',r,r')= -\frac{(\tau-\tau')}{16 \pi ^2 r r'\beta}\\ \times\left\{e^{\aleph_\beta} \left[2
i \text{Ci}\left(i\aleph_\beta\right)+2 \text{Si}\left(i\aleph_\beta\right)+\pi \right]\right.\\
\left.-e^{\aleph_\beta^*} \left[2 i \text{Ci}\left(i\aleph_\beta^*\right)-2 \text{Si}\left(i\aleph_\beta^*\right)+\pi \right]-2i\frac{\aleph_\beta-\aleph_\beta^*}{\aleph_\beta\aleph_\beta^*}\right\},\label{eq:greenminimal2}
\end{multline}
where
\begin{equation}
    \aleph_\beta\equiv\aleph_\beta(\tau,\tau',r,r')=\frac{ (r+r')+i(\tau -\tau ')}{\beta }.
\end{equation}

Taking the coincidence limit of the coordinates on $G_E^\beta(x,x')$ we have the contribution to the quadratic mean value of the field, which added to the one found in~\cite{mazzitelli91} returns the total value up to first order in $\eta^2$, i.e.,
\begin{multline}
    \braket{\hat{\Psi}^2}^{\xi=0}\equiv \braket{\hat{\Psi}^2}^{ML}+\braket{\hat{\Psi}^2}^{\beta}\\
    =-\frac{1}{4\pi^2r^2}\left \{ \eta^2\left [\frac{p}{2\sqrt{2}}-\frac{1}{6}\log\mu r  \right ]+e^{\frac{2r}{\beta}}\text{Ei}\left ( -\frac{2r}{\beta} \right ) \right \}, \label{eq:minimean}
\end{multline}
where $p=-0.39$, $\mu$ is a mass scale introduced by Mazzitelli and Lousto in the renormalization procedure, and $\text{Ei}$ is the exponential integral function. What is interesting about the mean value~\eqref{eq:minimean} is the appearance of a contribution in zeroth order of $\eta^2$ due exclusively to the boundary conditions at the singularity. In fact, it is direct to check that the new term we obtained vanishes as we take the limit $\beta\rightarrow 0$. Moreover, the first order term of our Green's function $G^\beta_{E,2}$ vanishes in the coincidence limit, so no contributions from the boundary conditions on the quadratic mean value can appear in order $\eta^2$.

\subsection{\label{sec:gmconform}Conformally coupled field}

In the case $\xi=1/6$, we must take $\nu_0=\frac{1}{2}\sqrt{1+\frac{4\eta^2}{3\alpha^2}}$ and apply it in Eq.~\eqref{eq:greenbeta}. Integration, however, becomes impractical to be done analytically, hence we will directly compute the series coefficients in~\eqref{eq:series}. After expanding the argument inside the integral in Eq.~\eqref{eq:greenbeta} in powers of $\eta^2$ and evaluating the integral to the first coefficient, we have
\begin{multline}
   G^{\beta}_{E,0}(\tau,\tau',r,r')= -\frac{i }{16 \pi ^2 r r'}\\ \times\left\{e^{\aleph_\beta} \left[-2
i \text{Ci}\left(-i\aleph_\beta\right)+2i \text{Shi}\left(\aleph_\beta\right)+\pi \right]\right.\\
\left.-e^{\aleph_\beta^*} \left[2 i \text{Ci}\left(i\aleph_\beta^*\right)-2i \text{Shi}\left(i\aleph_\beta^*\right)+\pi \right]\right\},\label{eq:greenconform0}
\end{multline}
where $\text{Shi}$ is the hyperbolic sine integral function. We were not able to evaluate analytically the first order term $G^\beta_{E,2}$.

Again the coincidence limit of the coordinates can be taken to obtain the quadratic mean value of the field, which to the lowest order of $\eta^2$ will be
\begin{equation}
    \braket{\hat{\Psi}^2}^{\xi=1/6}=-\frac{1}{4\pi^2r^2}e^{\frac{2r}{\beta}}\text{Ei}\left ( -\frac{2r}{\beta} \right )+\mathcal{O}(\eta^2). \label{eq:psisqzero}
\end{equation}
Like in the minimally coupled case, a contribution appears in order $\mathcal{O}(\eta^0)$. In particular, $\braket{\hat{\Psi}^2}^{\xi=1/6}$ and $\braket{\hat{\Psi}^2}^{\xi=0}$ are identical in the lowest order of $\eta^2$, which indicates a fluctuation independent of the curvature. Furthermore, for all values of the coupling constant within the range $0\leq\xi\ll 1/\eta^2$, the contribution in order $\mathcal{O}(\eta^0)$ remains the same as in Eq.~\eqref{eq:psisqzero}.

\section{\label{sec:back}Contributions to $\braket{\hat{T}_{\mu\nu}}$}

One of the most relevant quantities in semiclassical approaches to gravity is the energy-momentum tensor of the quantum fields propagating over the classical background of General Relativity. In this context, the computation of the energy-momentum tensor demands renormalization procedures to be followed. As discussed in~\cite{birrell1984quantum}, the renormalized v.e.v. of the energy-momentum tensor, $\braket{\hat{T}_{\mu\nu}}$, can be calculated in terms of the renormalized Green's function, as follows
\begin{equation}
    \braket{\hat{T}_\mu^{~\nu}(x)}_\text{ren}=\lim_{x'\rightarrow x}\left (\mathcal{D}_\mu^{~\nu}(x,x')G_\text{ren}(x,x')  \right ),\label{eq:emtdiff}
\end{equation}
where $\mathcal{D}_{\mu\nu}(x,x')$ is a non-local differential operator. One can find it by constructing the operator $\hat{T}_{\mu\nu}$ from the Lagrangian density of a scalar field and taking its expectation value w.r.t. a vacuum state~\cite{birrell1984quantum}.

The computation of this quantity can be highly non trivial since it is locally defined and we are operating over non-local objects (Green's functions) to obtain it. For instance, what Mazzitalli and Lousto do in~\cite{mazzitelli91} is to not explicitly calculate $\braket{\hat{T}_{\mu\nu}}$, but to use symmetry arguments and intrinsic properties of the definition of the tensor to at least find its form, which is
\begin{equation}
    \braket{\hat{T}_\mu^{~\nu}(x)}_\text{ren}^{ML}=\frac{1}{16\pi^2 r^4}\left [ A_\mu^{~\nu}(\xi,\eta^2)+B_\mu^{~\nu}(\xi,\eta^2) \log(\mu r) \right ],
\end{equation}
where $A_\mu^{~\nu}$ and $B_\mu^{~\nu}$ are both tensors with numerical components depending on the coupling constant $\xi$ in order $\mathcal{O}(\eta^2)$. The authors obtain explicitly only $B_\mu^{~\nu}$ as a consequence of the renormalization procedure. To find $A_\mu^{~\nu}$ it would be necessary to compute at least one of the components of $\braket{\hat{T}_\mu^{~\nu}}_\text{ren}^{ML}$ and to know its trace.

We are interested in the contributions to $\braket{\hat{T}_\mu^{~\nu}}_\text{ren}$ arising from the boundary conditions we have prescribed. In particular, we will compute the contributions to $\braket{\hat{T}_t^{~t}}$ which represents the energy density of the fields gravitating around the Monopole. To do so, we followed direct differentiation of the Green's function using Eq.~\eqref{eq:emtdiff}. The functional form of the operator $\mathcal{D}_t^{~t}(x,x')$ depends on the coupling parameter $\xi$, hence we will consider the minimally and conformally coupled cases separately. Only $\ell=0$ modes are influenced by the boundary conditions, its spherical symmetry implies the Green's function and, accordingly, the differential operator will depend only on $\tau~(\tau')$ and $r~(r')$. As we showed in Sec.~\ref{sec:gm}, in both cases the Green's functions are finite so no further renormalization procedures must be followed.

\subsection{\label{sec:emtmin}Minimally coupled field}

The differential operator in this case will be
\begin{equation}
  \mathcal{D}_t^{~t}(x,x')=-\frac{1}{2}\left ( \alpha^{-2}\frac{\partial^2 }{\partial \tau^2}+\alpha^2\frac{\partial^2 }{\partial r \partial r'} \right )
\end{equation}
and we must apply it on the Euclidean Green's function in Eq.~\eqref{eq:greenminimal}. After that, we take the coincidence limit, which leads to the $00-$th component of the energy-momentum tensor (energy density)
\begin{equation}
    \braket{\hat{T}_t^{~t}(x)}_\beta^{\xi=0}=-\frac{1-\eta^2}{8\pi^2r^4}\left [ 1+e^{\frac{2r}{\beta}} \left ( \frac{2r}{\beta}-1 \right )\text{Ei}\left ( -\frac{2r}{\beta} \right )\right ].\label{eq:emtmini}
\end{equation}
Again, in the limit $\beta\rightarrow 0$ the energy density vanishes, and the whole contribution will come from the Dirichlet boundary condition term, $\braket{\hat{T}_\mu^{~\nu}(x)}_\text{ren}^{ML}$. On the other hand, for the Neumann boundary condition ($\beta\rightarrow\infty$), the energy density diverges negatively to infinity, i.e., $\braket{\hat{T}_t^{~t}(x)}_\beta^{\xi=0}\rightarrow-\infty$.

\subsection{\label{sec:emtconform}Conformally coupled field}

In this case, the differential operator will be as follows
\begin{equation}
\mathcal{D}_t^{~t}(x,x')=-\frac{1}{6}\left ( 5\alpha^{-2}\frac{\partial^2 }{\partial \tau^2}+\alpha^2\frac{\partial^2 }{\partial r \partial r'} +\frac{\eta^2}{3r^2}\right ).
\end{equation}
As we only have available the Green's function in zeroth order for the conformally coupled field, we will follow the same procedure as before, i.e., to expand the energy-momentum tensor in a series of $\eta^2$. We will compute here only the contribution to the energy-momentum tensor in the lowest order of $\eta^2$ applying the differential operator over the Euclidean Green's function in Eq.~\eqref{eq:greenconform0}, which yields 
\begin{multline}
    \braket{\hat{T}_t^{~t}(x)}_\beta^{\xi=1/6}=-\frac{1}{8\pi^2r^4}\left [ \frac{2r}{3\beta}\right.\\
    \left.+\frac{e^{\frac{2r}{\beta}}}{3} \left ( \frac{4r^2}{\beta^2}+\frac{2r}{\beta}-1 \right )\text{Ei}\left ( -\frac{2r}{\beta} \right )\right ]\\
    +\mathcal{O}(\eta^2).\label{eq:emtconform}
\end{multline}
Similarly to the minimally coupled case, Dirichlet boundary condition leads to a vanishing contribution, and the Neumann one contibutes with a negatively divergent energy density.

\section{\label{sec:concl}Final Remarks}

The global monopole spacetime is characterized by a solid deficit angle proportional to $\eta^2$. This is responsible for the emergence of a naked singularity and the appearance of a strong curvature which is also proportional to $\eta^2$. As a matter of fact, we should expect only physical effects proportional to $\eta^2$ to appear. 

In this paper, however, we found that the nontrivial interaction between the quantum field and the classical singularity may bring contributions to the vacuum fluctuations which do not disappear in the limit $\eta\to0$. Only in the case where the field does not effectively realize the presence of the naked singularity (Dirichlet boundary condition) we have zero contribution in Minkowski limit. In this case, the fluctuations are purely topological. When nontrivial interactions are taken into account, an analytic contribution arises. Such results stand for a range of positive values of $\xi$ much smaller than a characteristic scale $1/\eta^2$. For negative values of $\xi$, the singularity may be perceived by non-spherically symmetric modes ($\ell=1,2,\dots$) as well, and boundary conditions might be necessary for them. For instance, such is the circumstance of the BTZ spacetime, which requires boundary conditions for all modes. However, we leave this case for further analysis in a future work.

The analytic contribution to the vacuum fluctuations seems to be originated by the local interaction between the field and the singularity, having nothing to do with the topology of spacetime. If nature abhors this behavior somehow, we have found out a preferred choice for the boundary conditions, namely the Dirichlet one. Otherwise, we have to accept such pathological behavior and try to find evidence in more realistic models, as the extreme Reissner-Norstrom spacetime.

\acknowledgments

It is a pleasure to acknowledge discussions with R.A. Mosna and A. Saa. The authors acknowledge support from FAPESP Grant No. 2013/09357-9. V.S. Barroso acknowledges finantial 
support from FAPESP Grant No. 2016/25963-4.
J.P.M.P. acknowledges support from FAPESP
Grant No. 2016/07057-6.

\end{document}